\documentclass[aps,prll,twocolumn,superscriptaddress,amsmath,amssymb]{revtex4}

\usepackage{natbib}
\usepackage{hyperref}
\usepackage{graphicx}
\usepackage{dcolumn}
\usepackage{bm}
\usepackage{hyperref}
\usepackage[mathlines]{lineno}
\usepackage[dvipsnames]{xcolor}

\begin{document}
	
\title{Putative hybridization gap in CaMn$_{2}$Bi$_{2}$ under applied pressure}

\author{M. M. Piva}
\email{mpiva@ifi.unicamp.br}
\affiliation{Instituto de F\'{\i}sica ``Gleb Wataghin'', UNICAMP, 13083-859, Campinas, SP, Brazil}
\affiliation{Los Alamos National Laboratory, Los Alamos, New Mexico 87545, USA}

\author{S. M. Thomas}
\affiliation{Los Alamos National Laboratory, Los Alamos, New Mexico 87545, USA}

\author{Z. Fisk}
\affiliation{Department of Physics and Astronomy, University of California, Irvine, California 92967}

\author{J.-X. Zhu}
\affiliation{Los Alamos National Laboratory, Los Alamos, New Mexico 87545, USA}

\author{J. D. Thompson}
\affiliation{Los Alamos National Laboratory, Los Alamos, New Mexico 87545, USA}

\author{P. G. Pagliuso}
\affiliation{Instituto de F\'{\i}sica ``Gleb Wataghin'', UNICAMP, 13083-859, Campinas, SP, Brazil}

\author{P. F. S. Rosa}
\affiliation{Los Alamos National Laboratory, Los Alamos, New Mexico 87545, USA}

\date{\today}
\begin{abstract}
	
	We report electrical transport measurements on CaMn$_{2}$Bi$_{2}$ single crystals under applied pressure. At ambient pressure and high temperatures, CaMn$_{2}$Bi$_{2}$ behaves as a single-band semimetal hosting N\'{e}el order at $T_{N}=150$~K. At low temperatures, multi-band behavior emerges along with an activated behavior typical of degenerate semiconductors. The activation gap is estimated to be $\Delta \sim 20$~K. Applied pressure not only favors the antiferromagnetic order at a rate of 0.40(2)~K/kbar, but also enhances the activation gap at $20$~kbar by about $70$~\%. This gap enhancement is typical of correlated narrow-gap semiconductors such as FeSi and Ce$_{3}$Bi$_{4}$Pt$_{3}$, and places CaMn$_{2}$Bi$_{2}$ as a Mn-based hybridization-gap semiconductor candidate. \textit{Ab initio} calculations based on the density functional theory are shown to be insufficient to describe the ground state of CaMn$_{2}$Bi$_{2}$.
	
\end{abstract}

\maketitle

\section{\label{sec:intro}Introduction}

Layered compounds with partially filled $d$- or $f$- shells often display emergent ground states, such as unconventional superconductivity, complex magnetic and electronic 
order, and non-Fermi-liquid behavior  \cite{review115, Paglione, Si}. A notable recent example is given by the iron-based superconductors, which crystallize in a layered 
tetragonal structure and host unconventional superconductivity, electronic nematicity, and different types of magnetism \cite{Rafael, Nema1, Nema2, Mn122, Pair}. Remarkably, when Fe is replaced with Mn, some of these materials crystallize in the hexagonal Ce$_{2}$SO$_{2}$-type structure with Mn-$Pn$ ($Pn =$ P, As, Sb, and Bi) layers that resemble the Fe-As layers. In fact, Mn sites in the hexagonal structure also display tetrahedral symmetry, as shown in Figure~\ref{fig1}(a), but the Mn layers form a puckered honeycomb lattice instead of a square net (Fig.~\ref{fig1}(b)).

In spite of these structural variations, the physical properties of the compounds $A$Mn$_{2}Pn_{2}$ ($A =$ Ca, Sr, Ba, and Eu), usually antiferromagnetic semiconductors, are generally affected by the pnictide 
size \cite{DC, Anand}. For instance, BaMn$_{2}Pn_{2}$ members crystallize in a tetragonal structure and show a trend of decreasing antiferromagnetic transition temperature ($T_{N}$) with increasing $Pn$ size, but no clear trend is seen in the energy gap ($\Delta$) \cite{BaP,BaMn2As2, DC, BaMn2Bi2}. By replacing Ba with Sr or Ca, these materials crystallize in the hexagonal structure mentioned above. SrMn$_{2}Pn_{2}$ compounds show no clear trend in $T_{N}$, but $\Delta$ increases with $Pn$ size \cite{BaP, CaSrMn2As2, DC}. 
Interestingly, the opposite trend is observed in CaMn$_{2}Pn_{2}$ members: $\Delta$ decreases with increasing $Pn$ size \cite{CaSrMn2As2, CaMn2Sb2}.

In particular, CaMn$_{2}$Bi$_{2}$ displays an antiferromagnetic order at $150$~K with magnetic moments lying in the honeycomb plane \cite{Cava}. 
At ambient pressure, CaMn$_{2}$Bi$_{2}$ has been reported to be a narrow-gap semiconductor with extremely large magnetoresistance \cite{Cava, MR}. At high temperatures, electrical resistivity 
measurements display metallic behavior, interpreted as the result of a strongly temperature dependent mobility \cite{Cava}. At low $T$, CaMn$_{2}$Bi$_{2}$ displays an activated behavior with $\Delta \sim  20$~K and a nonlinear Hall resistivity. Band-structure calculations suggest that one of the 3$d^{5}$ Mn electrons strongly hybridizes with the Bi $p$ bands giving rise to a hybridization gap. The other $d$ electrons remain localized giving rise to an ordered moment of 3.85~$\mu_{B}$ at $5$~K \cite{Cava}. In tetrahedral symmetry, Mn$^{2+}$ ions in a high-spin state could in fact support this scenario with one $t_{2g}$ orbital ($xy$) being more susceptible to hybridization. Moreover, angle-resolved photoemission spectroscopy (ARPES) experiments observed Mn-pnictide hybridization in BaMn$_{2}$As$_{2}$ and BaMn$_{2}$Sb$_{2}$ \cite{ARPES}. 

Applied pressure presents an ideal tuning parameter to test this framework. Typically, application of external pressure increases orbital overlap without introducing disorder in the system. If the small transport gap in CaMn$_{2}$Bi$_{2}$ were due to band structure of uncorrelated electrons, pressure thus should promote a metallic state \cite{McMahan}. However, if CaMn$_{2}$Bi$_{2}$ is indeed a hybridization gap semiconductor, the band gap could increase as a function of pressure as the hybridization between Mn 3$d$ and conduction electrons increases, akin to Ce$_{3}$Bi$_{4}$Pt$_{3}$ and FeSi under pressure \cite{Hundley, pressureCe343, FeSi}. In contrast, if the Mn 3$d$ ions become more localized with increasing pressure, the hybridization gap can decrease as a function of pressure similar to SmB$_{6}$ under pressure \cite{SmB6}. Therefore, electrical transport experiments under pressure are useful to answer whether CaMn$_{2}$Bi$_{2}$ is a hybridization gap semiconductor and to shed light on the evolution of its antiferromagnetic order and semiconducting gap.

Here we report electrical and Hall resistivity experiments under pressures to 20~kbar.  Our data show that $T_{N}$ increases at a rate of 0.4~K/kbar and the semiconducting gap is also enhanced with applied pressure, reaching $\sim$ 40~K at 20~kbar. Our results support the scenario in which CaMn$_{2}$Bi$_{2}$ is a hybridization gap semiconductor candidate. \textit{Ab initio} calculations, however, cannot describe the ground state of CaMn$_{2}$Bi$_{2}$.

\section{\label{sec:experiment}Experimental and Computational Details}

\begin{figure}[!t]
	\includegraphics[width=0.5\textwidth]{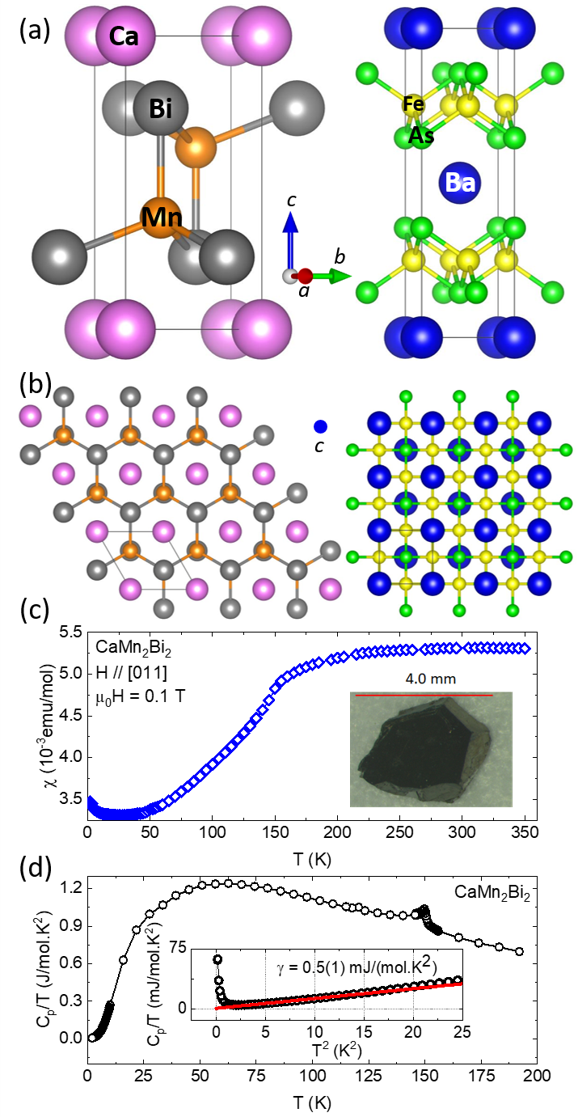}
	\caption{(a) Comparison between the crystalline structure of CaMn$_{2}$Bi$_{2}$ and BaFe$_{2}$As$_{2}$. (b) $c$-axis view of the crystalline structure of CaMn$_{2}$Bi$_{2}$ and BaFe$_{2}$As$_{2}$. (c) Magnetic susceptibility as a function of temperature. The inset shows a picture of a typical single crystal. (d) $C_{p}/T$ as a function of temperature. The inset shows $C_{p}/T$ as a function of $T^{2}$, at low temperatures.}
	\label{fig1}
\end{figure}

Single crystals of CaMn$_{2}$Bi$_{2}$ with typical size 2~mm x 1~mm x 1~mm were synthesized by the metallic flux technique with starting compositions Ca:Mn:Bi = 1.1:2:8. The mixture of elements was heated in a quartz tube under vacuum to 1100$^{\circ}$C at 75$^{\circ}$C/h, kept there for 4 hours and then cooled at 10$^{\circ}$C/h to 500$^{\circ}$C, at which temperature the tube was centrifuged to remove the Bi excess. The synthesized phase was checked by x-ray diffraction and energy-dispersive x-ray spectroscopy (EDX) resulting in lattice parameters $a$ = 4.636(2)~\AA \- and $c$ = 7.635(4)~\AA \ and 1.0(1):2.0(1):2.0(1) stoichiometry, respectively. A picture of a typical single crystal is displayed in the inset of Fig.~\ref{fig1}(c). Magnetic susceptibility was measured in a commercial Quantum Design MPMS. The specific heat was measured using a commercial Quantum Design PPMS small mass calorimeter that employs a 
quasi-adiabatic thermal relaxation technique. Electrical ($\rho_{xx}$) and Hall resistivity ($\rho_{xy}$) were measured in the (011) plane in their standard four-probe configurations. The crystal orientation was determined by a rocking curve measurement using a Cu K$_{\alpha}$ x-ray diffractometer.  All transport experiments were performed in a commercial Quantum Design PPMS with magnetic fields applied along the [011] direction and perpendicular to the current. Pressures up to 20~kbar were generated in a self-contained double-layer piston-cylinder-type Cu-Be pressure cell with an inner-cylinder of hardened NiCrAl. Daphne oil was used as the pressure transmitting medium and lead as a manometer.

The band structure of CaMn$_{2}$Bi$_{2}$ was determined from \textit{ab initio} calculations based on the density functional theory by using the plane-wave basis set and the projector augmented-wave method \cite{Blochl} as implemented in the Vienna simulation package (VASP) code \cite{Kresse}.  We applied both the standard spin-polarized generalized gradient approximation exchange-correlation functional according to the Perdew-Burke-Ernzerhof (PBE) \cite{Perdew} parametrization, and a modern nonlocal, range-separated, screened Coulomb potential hybrid density functional as proposed by Heyd, Scuseria, and Ernzerhof (HSE06) \cite{Heyd}. Spin-orbit coupling was included and a 500~eV energy cut-off was used to ensure the convergence of the total energy to 0.1 meV. The experimentally determined antiferromagnetic state was also considered and the Brillouin zone was sampled with 136 k-points in the irreducible wedge.

\section{\label{sec:results}Results and Discussion}

Figure \ref{fig1}(c) displays the magnetic susceptibility, $\chi(T)$, of CaMn$_{2}$Bi$_{2}$ as a function of temperature at $\mu_{0}H = 0.1$~T. At high temperatures, $\chi(T)$ is virtually temperature independent, as typically found in similar compounds with strong antiferromagnetic correlations \cite{CaSrMn2As2, McNally}. An inflection at 150~K marks $T_{N}$ followed by a small Curie tail at low temperatures suggesting the presence of a small amount of impurities. Figure \ref{fig1}(d) displays the specific heat, $C_{p}(T)/T$, as a function of temperature. A clear peak at high temperatures confirms $T_{N} = 150~K$. At low temperatures, the extrapolation of $C_{p}(T)/T$ $vs$ $T^{2}$ to $T = 0$~K provides an electronic coefficient of $\gamma = 0.5(1)$~mJ/mol.K$^{2}$. Although this coefficient is expected to be zero in an intrinsic semiconductor, a finite $\gamma$ may be observed when donor or acceptor levels are present. In fact, the prototypical hybridization gap semiconductor Ce$_{3}$Bi$_{4}$Pt$_{3}$ displays $\gamma = 3.3$~mJ/mol.K$^{2}$ \cite{Hundley}.

\begin{figure}[!t]
	\includegraphics[width=0.48\textwidth]{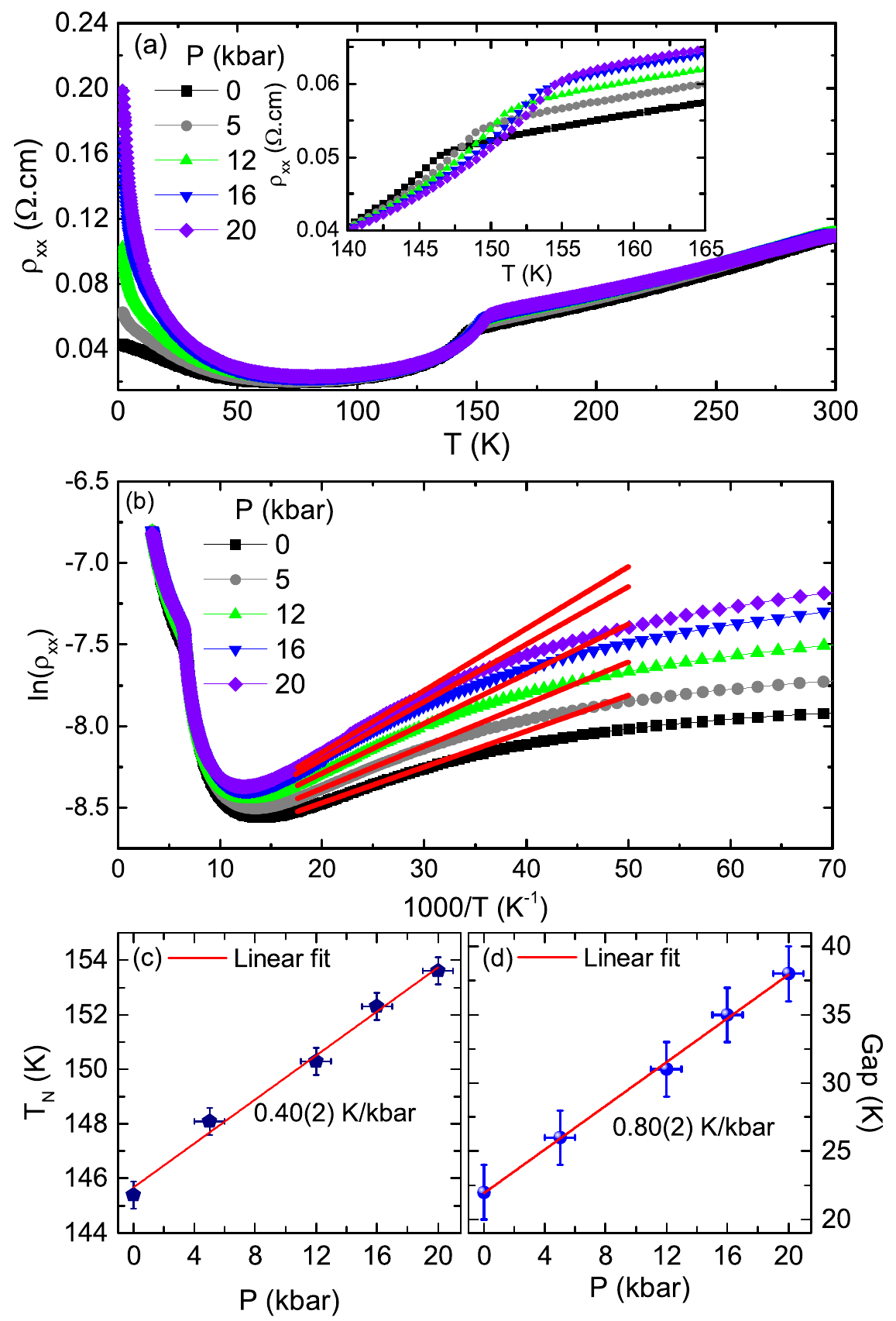}
	\caption{(a) $\rho_{xx}(T)$ for several pressures. The inset shows the antiferromagnetic transition. (b) Natural logarithm of $\rho_{xx}$ as a function of $1/T$. The red lines are the extrapolation of linear fits. Panels (c) and (d) show the pressure evolution of $T_{N}$ and $\Delta$, respectively.}
	\label{RxT}
\end{figure}

Figure \ref{RxT}(a) shows the temperature dependence of the electrical resistivity, $\rho_{xx}(T)$, at different applied pressures. At high temperatures, $\rho_{xx}(T)$ decreases with decreasing temperature, which was previously associated with an increase in the mobility rather than a metallic behavior \cite{Cava}. However, these data could be also explained by a semimetallic picture. As we will discuss later, the carrier density at high temperatures is $\sim 10^{18}$~holes/cm$^{3}$, typical of semimetals. Moreover, a decrease in the Hall coefficient with temperature is also consistent with a semimetal behavior, bismuth being a classic example \cite{Bismuth}.

\begin{figure}[!t]
	\includegraphics[width=0.48\textwidth]{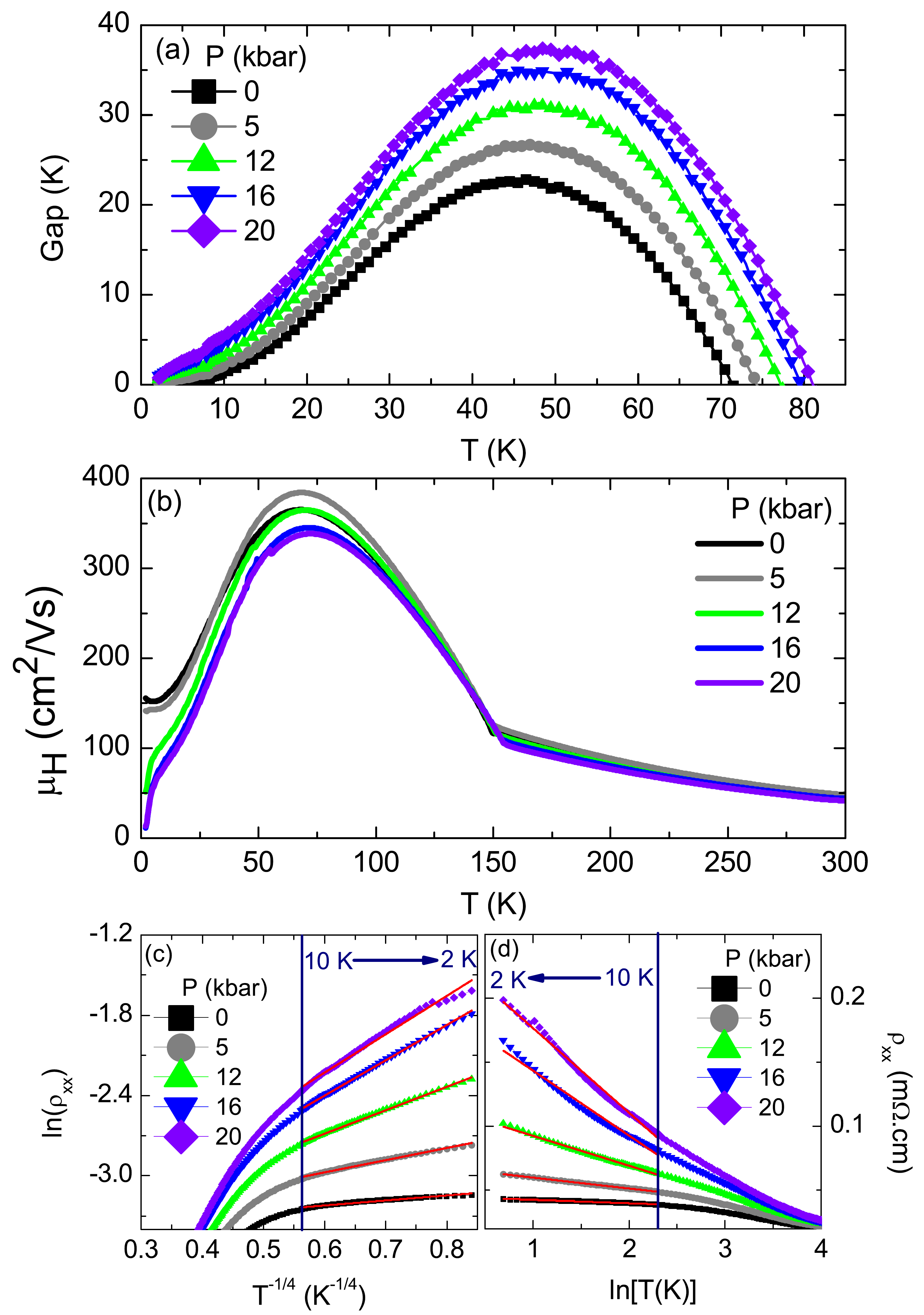}
	\caption{(a) $\Delta (T)$ for several pressures. (b) Carrier mobility as a function of temperature. (c) ln$(\rho_{xx})$ as a function of $T^{-1/4}$. (d) $\rho_{xx}$ as a function of ln$(T)$.}
	\label{GapxT}
\end{figure} 

As temperature is further decreased, a kink in $\rho_{xx}(T)$
signals the onset of $T_{N}$ at around 145~K (inset of Fig.~\ref{RxT}(a)). Below $T_{N}$, $\rho_{xx}(T)$ decreases rapidly, likely due to the reduced magnetic scattering in the ordered state.
Under applied pressure, $T_{N}$ increases, reaching a maximum of 154~K at 20~kbar. We note that, at ambient pressure, $T_{N}$ is slightly lower than $T_{N} = 150$~K  previously reported \cite{Cava, MR} due to an essentially pressure-independent temperature gradient between the sample and the thermometer. Our crystals measured outside the pressure cell show similar $T_{N}$ values as in Refs. \cite{Cava, MR}. 

At low temperatures and ambient pressure ($T \leqslant 65$~K), $\rho_{xx}(T)$ increases with decreasing temperature. Applying pressure to CaMn$_{2}$Bi$_{2}$ enhances this semiconducting-like behavior. As a result, the activation gap extracted from linear fits (from 35~K to 55~K) of an Arrhenius plot, ln$(\rho_{xx}) = $ln$(\rho_{0}) + \Delta/T$, displays a linear increase as a function of pressure (Fig.~\ref{RxT}(b)). Figures~\ref{RxT}(c) and (d) display a summary of the behavior of $T_{N}$ and $\Delta$ as a function of pressure, respectively. The $T-P$ phase diagram obtained from our data reveals an increase in $T_{N}$ as a function of pressure at a rate of 0.40(2)~K/kbar. Moreover, the activation energy ($\Delta$) increases from $\sim$ 20~K to $\sim$ 40~K at 20~kbar with a slope of 0.80(2)~K/kbar.

We note, however, that the gap value is an estimation and should be taken with caution. First, $T$ is greater than $\Delta/k_{B}$  in the temperature range used for the fits and, therefore, the apparent value of the gap cannot be taken at face value. Further, Fig.~\ref{GapxT}(a) presents the activation energy as a function of temperature, which was extracted by taking the first derivative of the data displayed in Fig.~\ref{RxT}(b). At higher temperatures, the extracted gap is temperature dependent and increases with decreasing temperature. This result may be an indication that the Fermi level is changing with temperature, characteristic of extrinsic semiconductors, or due to the changes in the lattice parameters caused by thermal contraction, as seen in Ce$_{3}$Bi$_{4}$Pt$_{3}$ \cite{Thermal}. Therefore, the gap extracted from our transport measurements may not be the intrinsic gap of the material, and spectroscopic measurements are necessary. 
Moreover, we extracted the carrier mobility ($\mu_{H}$) by performing $\rho_{xy}$ measurements as a function of temperature with an applied field of 5~T and then by dividing it by $\rho_{xx}$ and 5~T. $\mu_{H}$ first increases below $T_{N}$ due to a reduction in spin-disorder scattering and then decreases with decreasing temperature (Fig.~\ref{GapxT}(b)), suggesting that impurity conduction is taking place \cite{Book} or that the localization and/or the Kondo scattering are increasing.

\begin{figure}[!t]
	\includegraphics[width=0.48\textwidth]{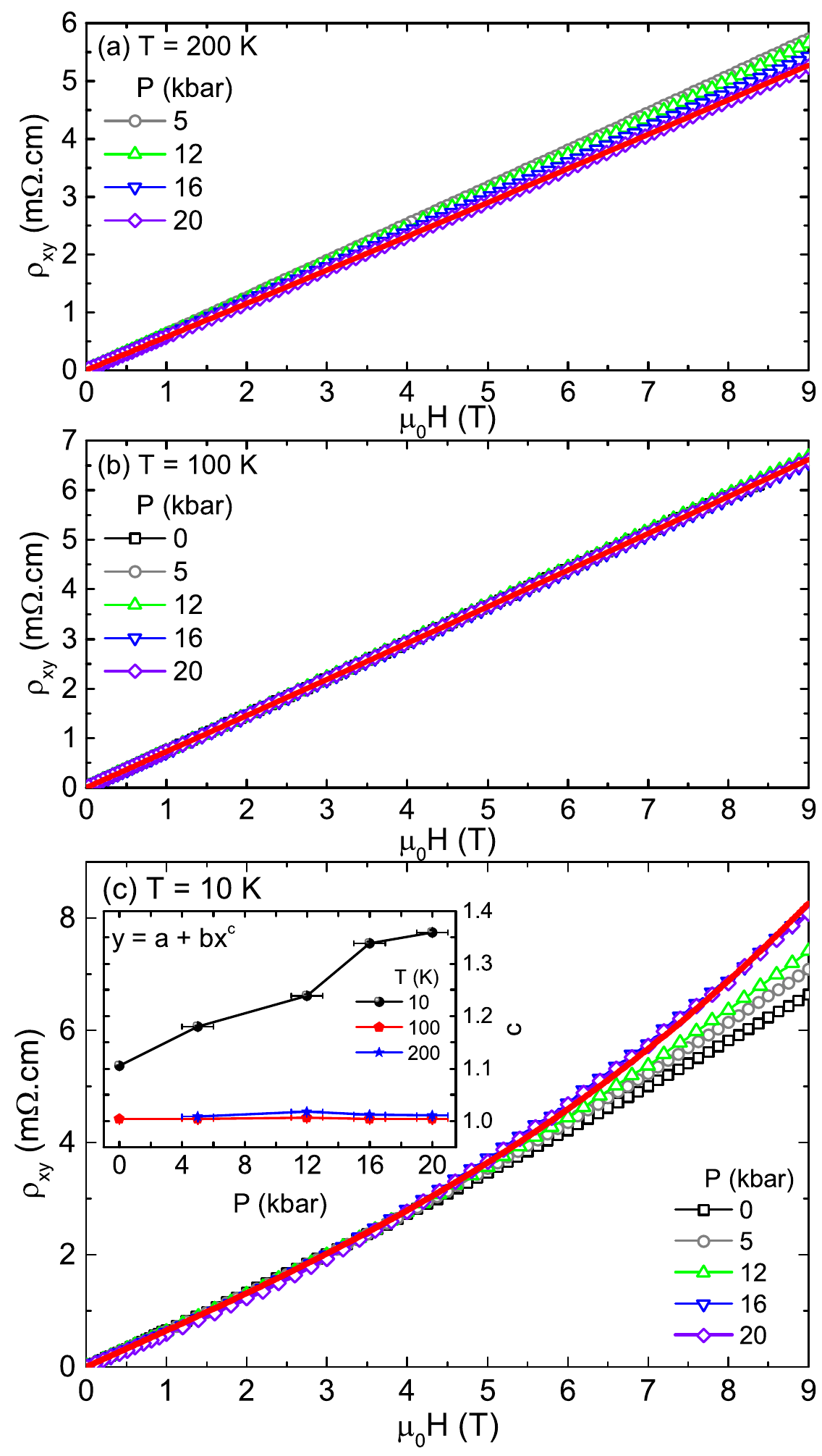}
	\caption{(a) $\rho_{xy}$ as a function of applied magnetic field for several pressures at 200~K (b) 100~K and (c) 10~K, the red lines are two-band model fits. The inset shows the $c$ exponents of allometric fits.}
	\label{HallxH}
\end{figure}

As a result, a possible scenario is the presence of variable-range hopping conduction, as disorder may play a significant role at low temperatures ($T \leqslant 20$~K) due to the presence of in-gap states \cite{CaMn2Sb2, Ingap2}. In fact, we can fit the resistance below 10~K to $\rho_{xx}(T) = \rho_{0}e^{(T_{0}/T)^{1/4}}$, as displayed in Fig.~\ref{GapxT}(c). We note that $T_{0}$ increases with pressure, even though a more metallic-like ground state is expected under applied pressure, as observed in the manganites LaMnO$_{3+\delta}$ and Nd$_{0.62}$Pb$_{0.38}$MnO$_{3-\delta}$ \cite{Manganites1, Manganites2}. This result indicates that the localization radius decreases with pressure in CaMn$_{2}$Bi$_{2}$. Another possibility is that, at low temperatures, Kondo scattering may start to develop, but it is prevented from percolating and forming a Kondo lattice due to the small carrier density in this compound. Indeed, our low temperature data can also be described by a $ln(T/T_{K})$ Kondo behavior for $\rho_{xx}$, as shown in Fig.~\ref{GapxT}(d). However, this is unlikely to be the case, as CaMn$_{2}$Bi$_{2}$ presents a very small Sommerfeld coefficient, which is not consistent with the presence of Kondo scattering. Nevertheless, the increase of the activation energy as a function of pressure in an intermediate temperature range, roughly from 35~K to 55~K, is similar to that of Ce$_{3}$Bi$_{4}$Pt$_{3}$ \cite{Ingap2}.  

\begin{figure*}[!t]
	\includegraphics[width=\textwidth]{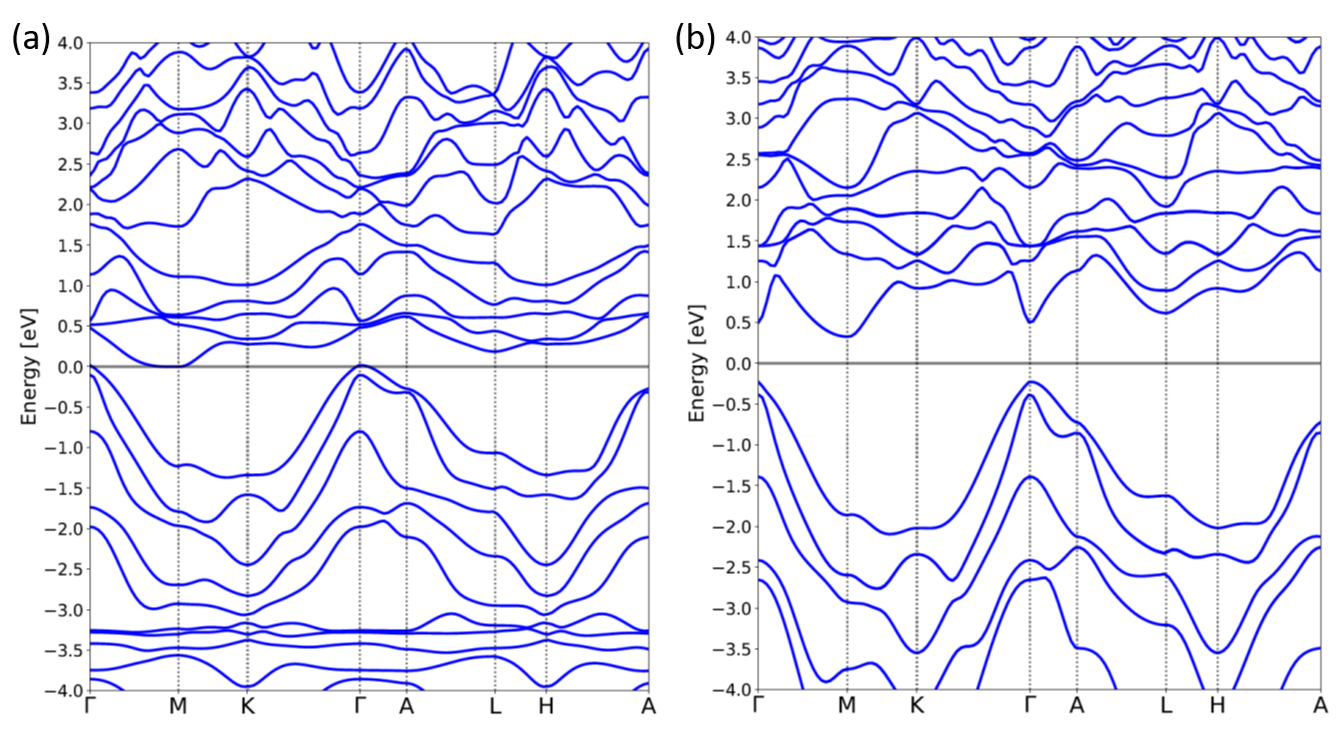}
	\caption{(a) DFT-PBE electronic band structure calculations. (b) Electronic band structure calculations considering the HSE06 functional.}
	\label{calc}
\end{figure*}

Figure \ref{HallxH} shows the Hall resistivity as a function of applied field for several pressures at three different temperatures. At high temperatures, $\rho_{xy}$ is linear with field at all pressures (Figs.~\ref{HallxH}(a) and (b)). This linear response indicates the presence of one dominant type of carrier (holes in this case) and that the compound can be treated as a single band system. By performing linear fits in this regime, we extracted a carrier density $n$ of $9.8(1)\times 10^{17}$~holes/cm$^{3}$ at 5~kbar and 200~K, which reaches $1.1(1)\times 10^{18}$~holes/cm$^{3}$ at 20~kbar. At 100~K, $n$ is nearly constant as a function of pressure with a value of $8.5(1)\times 10^{17}$~holes/cm$^{3}$.  
At $T = 10$~K, however, the Hall resistivity is nonlinear already at ambient pressure, and this non-linearity is enhanced by the application of external pressure (Fig.~\ref{HallxH}(c)). To better visualize the evolution of the Hall resistivity curvature as a function of pressure, we have performed allometric fits ($y = a + bx^{c}$) to all curves. The inset of Fig.~\ref{HallxH}(c) shows the value of the exponent $c$ as a function of pressure, which confirms the pressure-independent linear behavior of the Hall resistivity at high temperatures, both above and below $T_{N}$.  At 10~K, however, applied pressure is responsible for an increase of the $c$ exponent by 25~$\%$ at 20~kbar.

The observed non-linearity in $\rho_{xy}$ suggests multi-band effects. In fact, our data can be fit to a two-band model,

\begin{equation}
\rho_{xy} (B) = \frac{B}{e}\frac{(n_{h}\mu_{h}^2 - n_{e}\mu_{e}^2) + (n_{h} - n_{e})\mu_{e}^2\mu_{h}^2 B^2}{(n_{h}\mu_{h} + n_{e}\mu_{e})+[(n_{h} - n_{e})\mu_{e}\mu_{h} B]^2}
\end{equation}

\noindent where $n$ and $\mu$ are the carrier density and the mobility, respectively, for holes (h) and electrons (e). Representative fits are shown as solid lines in Fig.~\ref{HallxH}. We note that this fit is under constrained. However, for the fit to converge, the carrier densities need to be nearly identical, regardless of the values chosen for the mobilities, suggesting that CaMn$_{2}$Bi$_{2}$ is nearly compensated at $T = 10$~K. We note that this compensation could be responsible for the extremely large magnetoresistance reported recently in Ref.~\cite{MR}. Nevertheless, we also note that these fits may not be unique and should be taken with caution, especially because there are no experimental constraints on the carrier densities and mobilities obtained from, for example, quantum oscillation measurements.

Finally, we discuss the band structure of CaMn$_{2}$Bi$_{2}$ from first-principles calculations.
As shown in Fig. 5(a), the obtained DFT-PBE band structure agrees with a previous calculation with a full-potential linearized augmented plane-wave method~\cite{Cava}. However, there is a slight energy overlap between valence and conduction bands, which suggests that CaMn$_{2}$Bi$_{2}$ presents a semimetallic behavior. As shown in Fig. 5(b), the HSE06 functional opens a semiconducting band gap of 0.5~eV (5800~K), which is much larger than the experimental extracted excitation gap ($\sim 20$~K). These discrepancies indicate that independent of the functional DFT is not the most appropriate approach to evaluate the ground-state of correlated CaMn$_{2}$Bi$_{2}$ and DFT+DMFT calculations are required. In fact, DFT+DMFT calculations in related Mn pnictides LaMnPO and BaMn$_{2}$As$_{2}$ reveal a large Hund's coupling ($\sim$ 1~eV) as well as a large on-site Coulomb repulsion ($\sim$ 8~eV) owing to the half-filled Mn $d$ shell. This suggests that 3$d^{5}$ Mn pnictides are more correlated than their iron 3$d^{6}$ analogs \cite{McNally}. Moreover, recent experiments on  CsFe$_{2}$As$_{2}$ and NaFe$_{1-x}$Cu$_{x}$As$_{2}$ indicate that a Fe 3$d^{5}$ configuration leads to a more correlated, insulating state that competes with superconductivity \cite{Eilers, Song}.    

In most materials, the application of external pressure tends to favor metallic behavior, increasing the hybridization between local states and conduction electrons \cite{McMahan}. Our results, however, show that the application of external pressure in CaMn$_{2}$Bi$_{2}$ has the opposite behavior, and the low temperature resistivity is enhanced. We argue that this behavior is consistent with CaMn$_{2}$Bi$_{2}$ being a hybridization gap semiconductor: the Mn 3$d$-conduction electron hybridization is enhanced by the application of external pressure, leading to a larger hybridization gap and favoring a more insulating-like behavior. This scenario agrees with a tetrahedral crystal field scheme in which the $t_{2g}$ $xy$ orbitals hybridize more strongly with conduction electrons. Polarized ARPES measurements would be valuable to provide a direct confirmation of this scenario. Microscopic experiments, such as nuclear magnetic resonance, would be useful to investigate the evolution of the Mn 3$d$ hybridization as a function of pressure. Finally, optical measurements will be key to directly probe the activated gap value.

\section{Conclusions}

We report electrical and Hall resistivity experiments in antiferromagnetic CaMn$_{2}$Bi$_{2}$ single crystals under pressures to 20~kbar. At high temperature, CaMn$_{2}$Bi$_{2}$ behaves as a single-band semimetal, whereas it displays a nonlinear response of the Hall resistivity at low temperatures. Our data suggest that CaMn$_{2}$Bi$_{2}$ is an extrinsic narrow-gap semiconductor with a putative activation gap of 20~K. The application of external pressure increases $T_{N}$ at a rate of 0.40(2)~K/kbar, and enhances the extracted gap to $\sim$ 40~K at 20~kbar, consistent with the picture of a hybridization driven gap. Furthermore, band structure calculations predict the presence of a 0.5~eV (5800~K) gap considering the HSE06 functional. The large discrepancy between the calculated gap and the experimentally observed one ($\sim$ 20~K) indicates that DFT is not suitable to evaluate the ground-state of CaMn$_{2}$Bi$_{2}$. Performing DFT+DMFT calculations may improve the agreement between experimental and theoretical results. 

\section{Acknowledgments}

We would like to acknowledge constructive discussions with C. Kurdak, F. Malte Grosche, F. Ronning, Yongkang Luo, and M. F. Hundley. This work was supported by the S\~ao Paulo Research Foundation (FAPESP) grants 2015/15665-3, 2017/25269-3, 2017/10581-1, CAPES and CNPq, Brazil. Work at Los Alamos was performed under the auspices of the U.S. Department of Energy, Office of Basic Energy Sciences, Division of Materials Science and Engineering.
P.~F.~S.~R. acknowledges support from the Laboratory Directed Research and Development program of Los Alamos National Laboratory under project number 20180618ECR. Scanning electron microscope and energy dispersive X-ray measurements were performed at the Center for Integrated Nanotechnologies, an Office of Science User Facility operated for the U.S. Department of Energy (DOE) Office of Science. The electronic structure part of the work was supported by the U.S. DOE BES E3B5.

\end{document}